\documentclass[fleqn,10pt]{wlscirep}
\usepackage{float}
\title{MUSAN: A Music, Speech, and Noise Corpus}

\author[1,*]{David Snyder}
\author[1]{Guoguo Chen}
\author[1]{Daniel Povey}
\affil[1]{Center for Language and Speech Processing, The Johns Hopkins University, Baltimore, MD 21218, USA}

\affil[*]{david.ryan.snyder@gmail.com}


\begin{abstract}
This report introduces a new corpus of music, speech, and noise. This dataset is
suitable for training models for voice activity detection (VAD) and music/speech discrimination. Our corpus is released under
a flexible Creative Commons license.
The dataset consists of music from several genres, speech from twelve languages, 
and a wide assortment of technical and non-technical noises.
We demonstrate use of this corpus for music/speech discrimination
on Broadcast news and VAD for speaker identification. 
\end{abstract}
\begin{document}

\flushbottom
\maketitle
\thispagestyle{empty}

\section{Introduction}

Classification of audio into speech, music, or other nonspeech categories has several practical applications. Voice activity detection (VAD)
is often an essential preprocessing step for other speech technologies, such 
as speech recognition, speaker diarization, or speaker verification. In some applications, on hold music needs to be detected and removed.

Gaussian mixture models (GMM) are a simple and popular model for VAD and music/speech discrimination
\cite{kenny, greg_sell}. In \cite{kenny}, several frame-level VAD techniques were compared for speaker verification, and an unsupervised
GMM-based VAD was found to produce the best results on NIST SRE 2008. We also make use of GMM-based
systems to demonstrate this corpus. 
Moreover, our focus is on providing
the data; therefore, we do not explore complex models or sophisticated features.

Most publicly available corpra for music and speech
discrimination that provide raw audio do not address the copyright of the data sources nor appear to have 
permission to redistribute the
data. For instance, the GTZAN Music/Speech dataset \cite{gtzan} is widely used, but it appears that 
permission was not given by the copyright holders 
to redistribute the work. Other corpra, such as the Million Song database \cite{million} circumvent intellectual 
property issues by providing only features. However, this limits the user to building systems based only on the provided feature type. In contrast, our corpus is compiled from Creative Commons and US Public Domain sources, so we are free to redistribute the original
audio.

\section{The corpus}

The corpus consists of approximately 109 hours of audio that
is in the US Public Domain or under a Creative Commons
license. The directories are partitioned into speech,
music, and noise and by data source (i.e., the website we downloaded the
content from). It is freely available at 
\href{http://www.openslr.org/resources.php}{OpenSLR}. All audio is formatted as 16kHz WAV files. 

Each subdirectory contains a LICENSE file which connects a
WAV file to its respective license and provides attribution appropriate to the license type. For example, a music
entry in a LICENSE file may contain the filename, title, artist,
a url to the source, and a summary of the license.  All files in this
corpus fall under a Creative Commons license or are considered to be in
the US Public Domain.  
To broaden the use of this corpus, we only include content that allows for commercial use. Please refer to the
\href{http://creativecommons.org/licenses/}{Creative Commons}
licenses for more information.

Most directories contain an ANNOTATIONS file which
provides potentially useful metadata. For example, music is annotated for the presence or absence
of vocals and by genre(s). The READMEs in each subdirectory describe the types of annotations in more detail.

\subsection{Speech}

This portion of the corpus consists of about
60 hours of speech. It contains 20 hours and 21 minutes of read speech from \href{https://librivox.org/}{Librivox},
all of which are in the Public Domain.
Each WAV file is an entire chapter of a book, read by one
speaker. Approximately half of the Librivox
recordings are in English, and the remainder
are from eleven other languages. Annotations for speaker and language 
are provided. The rest of the speech portion
consists of 40 hours and 1 minute of US government 
(federal and various states) hearings, committees and debates that are believed to be in the US Public Domain. 
These files have been obtained from the \href{https://archive.org/}{Internet Archive} and the \href{http://s1.sos.mo.gov/records/missourichannel/senate}{Missouri Channel} senate archives.
These recordings are entirely in English.

\subsection{Music}

The music portion of the corpus has been downloaded from \href{https://www.jamendo.com/en/welcome}{Jamendo}, \href{http://freemusicarchive.org/}{Free Music Archive},
\href{http://incompetech.com/music/royalty-free/collections.php}{Incompetech}, and \href{http://www.hdclassicalmusic.com/}{HD Classical Music}. It is 42 hours and 31 minutes. The music is divided into Western art music (e.g.,
Baroque, Romantic, and Classical) and popular genres (e.g., jazz, bluegrass, hiphop, etc). Annotations for genre, artist,
and the presence or absence of vocals are provided. 
For Western art music, the composer is also
provided. These files have all been released under some form of the Creative Commons license.
Care was taken to ensure that any of these files can be used for commercial purposes.

\subsection{Noise}

This portion of the corpus contains 929 files of assorted noises, with a total duration of about 6 hours. These range from technical
noises, such as DTMF tones, dialtones, fax machine noises, and more, as well as ambient
sounds, such as car idling, thunder, wind, footsteps, paper rustling, rain, animal noises, etc.
We do not include recordings with intelligible speech. However, some recordings are crowd noises with indistinct voices. The files were downloaded from \href{https://www.freesound.org/}{Free Sound} and \href{http://soundbible.com/}{Sound Bible}.
The Free Sound part of the corpus is
Public Domain material and the Sound Bible
part is CC licensed.

\section{Experiments}

We demonstrate use of this corpus by training several simple systems for music/speech discrimination and voice activity detection. Our experiments use the Kaldi ASR toolkit \cite{kaldi}.

\subsection{Music/speech discrimination}
\label{sec:music_speech}
We train two GMMs on the speech and music portions of the corpus. The features
are 20 MFCCs with sliding window mean normalization. The first through fourth order deltas are appended to the MFCCs. We compare with an identical system trained on the GTZAN
dataset.

We test the systems on Broadcast news \cite{hub4}. This evaluation is similar to \cite{greg_sell}. The test audio is
separated into speech and music segments. 
Any overlapping segments were excluded from the evaluation. Classification is performed on the entire segment, by taking the majority of the frame-level decisions. We evaluate at the
equal error-rate (EER) operating point.

\subsection{Music/speech discrimination results}
\begin{table}[!ht]
\begin{center}
\begin{tabular}{l|cccccc}
\hline
EER(\%)/$K$  & 4 & 8 & 16 & 32 & 64 & 128 \\ \hline \hline
MUSAN   & 4.43 & 3.85 & 3.75 & 3.85 & 3.95 & 4.14 \\
GTZAN   & 3.85 & 3.95 & 3.85 & 4.05 & 4.05 & 3.95 \\ \hline
\end{tabular}
\end{center}
\caption{A comparison of systems trained on the MUSAN and GTZAN corpra. We calculate the EER(\%) using GMMs with $K \in \{4,8,16,32,64,128\}$ components. }
\label{bn}
\end{table}

We experiment with GMMs of varying sizes. In Table \ref{bn} we see that, overall, performance isn't affected much by the GMM size, and similar performance is achieved with the models trained on the GTZAN and MUSAN corpra. 

\subsection{Speaker recognition}

\begin{figure}[th]
\centerline{\includegraphics[width=10 cm]{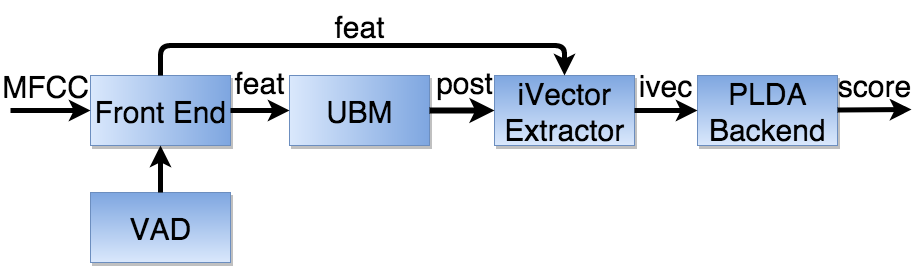}}
\caption{Speaker recognition schema.}
\label{fig:schema}
\end{figure}

The speaker recognition system is a typical i-vector-based system. It uses a GMM universal background
model (UBM) and a PLDA backend (See Figure \ref{fig:schema}). It is based on the GMM system described in \cite{supgmm}. The raw features are 20 MFCCs with a 25ms frame-length. The front end performs a sliding window mean-normalization over a 3 second window, concatenates delta and acceleration, and uses the voice activity detection (VAD) decisions to remove unvoiced frames.

Our evaluation is based on the core NIST SRE 2010 \cite{sre10} evaluation. To better expose the effects of voice activity detection, we modify the original evaluation by restricting the amount of speech available at test time. In the original evaluation, the utterances are all approximately 5 minutes long. In this test, only the first $n$ seconds of speech are available to the system at test time. This simulates a practical scenario in which recognition needs to be performed quickly at test time, and better exposes the benefits of a more accurate VAD decisions.

\subsubsection{Energy VAD}
\label{sec:energy}

Our baseline system uses a simple energy VAD. The
energy VAD classifies a frame as speech or nonspeech by 
using information about the average log-energy in a given
window, centered around the current frame.

\subsubsection{GMM and Energy VAD}
This system uses a GMM-based VAD trained on this corpus in addition to the energy-based VAD
described in Section \ref{sec:energy}.

Three full-covariance GMMs, each with 8 components, are trained on the speech, music, and noise portions of the corpus. We use only the music without vocals, and only the Librivox portion of the speech data. The features are identical to those used for music/speech discrimination in Section \ref{sec:music_speech}.

To classify a frame a speech or nonspeech, we first apply the energy-based VAD. All of the frames classified as speech are then refined by the GMM-based VAD. We reclassify a frame by selecting the GMM with the highest posterior. The priors are 0.07 for music, 0.75 for speech, and 0.18 for noise, and were selected by tuning on an out-of-domain dataset. Frames classified as music or noise are mapped to nonspeech.

\begin{table}[!ht]
\begin{center}
\begin{tabular}{l|ccccccc}
\hline
VAD/EER(\%)  & Max & 60s & 10s & 5s & 3s & 2s & 1s \\ \hline \hline
Energy       & 2.54 & 2.99 & 6.52 & 9.98 & 14.17 & 17.48 & 23.10 \\
GMM+Energy   & 2.37 & 2.65 & 5.01 & 8.06 & 11.23 & 14.50 & 20.25 \\ \hline
Rel. Improv. & 6.70 & 11.37 & 23.16 & 19.24 & 20.75 & 17.05 & 12.34 \\ \hline
\end{tabular}
\end{center}
\caption{Speaker recognition performance on the core NIST SRE 2010 evaluation with and without the GMM-based VAD.}
\label{gender_ind}
\end{table}

\subsection{VAD Results}

In Table \ref{gender_ind} we see that the addition of a GMM-based VAD
improves performance on the speaker recognition task. The speaker verification results are
improved for all amounts of speech, but the benefit is greater when less speech is available.

\section{Acknowledgments}
This material is based upon work partially supported by the National Science Foundation Graduate Research Fellowship under Grant No. 1232825. Any opinion, findings, and conclusions or recommendations expressed in this material are those of the authors(s) and do not necessarily reflect the views of the National Science Foundation.

This work was partially supported by
Spoken Communications. 

\section{Conclusion}
We introduced a corpus for training music/speech discrimination and voice activity detection (VAD) systems. This corpus was
collected entirely from US Public Domain
and Creative Commons sources, and provides the
raw audio. We showed that this 
dataset can be used to train simple classifiers for music/speech discrimination and also for frame-level VAD.

\bibliographystyle{abbrv}
\bibliography{mybib}

\begin{thebibliography}{1}

\bibitem{sre10}
The {NIST} year 2010 speaker recognition evaluation plan.
\newblock \url{http://www.itl.nist.gov/iad/mig/tests/sre/2010/}, 2010.

\bibitem{kenny}
J.~Alam, P.~Kenny, P.~Ouellet, T.~Stafylakis, and P.~Dumouchel.
\newblock Supervised/unsupervised voice activity detectors for text-dependent
  speaker recognition on the rsr2015 corpus.
\newblock In {\em Proc of the Odyssey speaker and language recognition
  workshop,(Odyssey’14)}, pages 123--30, 2014.

\bibitem{million}
T.~Bertin-Mahieux, D.~P. Ellis, B.~Whitman, and P.~Lamere.
\newblock The million song dataset.
\newblock In {\em {Proceedings of the 12th International Conference on Music
  Information Retrieval ({ISMIR} 2011)}}, 2011.

\bibitem{hub4}
J.~F. W.~F. David~Graff, John~Garofolo and D.~Pallett.
\newblock 1996 english broadcast news speech (hub4).
\newblock Linguistic Data Consortium, Philadelphia, 1997.

\bibitem{supgmm}
D.~P. David~Snyder, Daniel Garcia-Romero.
\newblock Time delay deep neural network-based universal background models for
  speaker recognition.
\newblock In {\em ASRU}, 2015.

\bibitem{kaldi}
D.~Povey, A.~Ghoshal, G.~Boulianne, L.~Burget, O.~Glembek, N.~Goel,
  M.~Hannemann, P.~Motl{\'\i}{\v{c}}ek, Y.~Qian, P.~Schwarz, et~al.
\newblock The {K}aldi speech recognition toolkit.
\newblock In {\em Proceedings of the Automatic Speech Recognition \&
  Understanding (ASRU) Workshop}, 2011.

\bibitem{greg_sell}
G.~Sell and P.~Clark.
\newblock Music tonality features for speech/music discrimination.
\newblock In {\em Acoustics, Speech and Signal Processing (ICASSP), 2014 IEEE
  International Conference on}, pages 2489--2493. IEEE, 2014.

\bibitem{gtzan}
G.~Tzanetakis.
\newblock Gtzan music/speech.
\newblock \url{http://marsyasweb.appspot.com/download/data_sets/}.
\newblock Accessed: 2015-11-24.

\end{thebibliography}
\end{document}